\documentclass[prl,aps,showpacs,lengthcheck,superscriptaddress,twocolumn]{revtex4}
\usepackage{amsmath}
\pdfoutput=1
\usepackage{xcolor}
\usepackage{amsfonts}
\usepackage{graphicx}
\usepackage{bm}
\usepackage{blindtext}

\newcommand{\beq}{\begin{equation}}
\newcommand{\eeq}{\end{equation}}
\newcommand{\beqa}{\begin{eqnarray}}
\newcommand{\eeqa}{\end{eqnarray}}

\begin{document}

\title{Aggregation and Segregation of Confined Active Particles}

\author{X.~Yang}
\affiliation{Physics Department, Syracuse University, Syracuse NY 13244, USA}
\author{M.~L. Manning}
\affiliation{Physics Department, Syracuse University, Syracuse NY 13244, USA}
\affiliation{Syracuse Biomaterials Institute, Syracuse University, Syracuse NY 13244, USA}
\author{M.~C. Marchetti}
\affiliation{Physics Department, Syracuse University, Syracuse NY 13244, USA}
\affiliation{Syracuse Biomaterials Institute, Syracuse University, Syracuse NY 13244, USA}

\begin{abstract}
We simulate a model  of self-propelled disks with soft repulsive interactions confined to a box in two dimensions. For small rotational diffusion rates, monodisperse disks spontaneously accumulate at the walls. At low densities, interaction forces between particles are strongly inhomogeneous, and a simple model predicts how these inhomogeneities alter the equation of state.  At higher densities, collective effects become important. We observe signatures of a jamming transition at a packing fraction $\phi \sim 0.88$, which is also the jamming point for non-active athermal monodisperse disks.  At this $\phi$, the system develops a critical finite active speed necessary for wall aggregation.  At packing fractions above $\phi \sim 0.6$, the pressure decreases with increasing density, suggesting that strong interactions between particles are affecting the equation of state well below the jamming transition.  A mixture of bidisperse disks segregates in the absence of any adhesion, identifying a new mechanism that could contribute to cell sorting in embryonic development.

\end{abstract}
\maketitle
Minimal models of self-propelled particles (SPP) have provided much insight into the emergent behavior of non-equilibrium, active systems where energy is injected at the scale of the individual constituents. This novel class of materials spans many length scales, ranging from bird flocks to bacterial swarms, cell layers and synthetic microswimmers~\cite{Marchetti2013}.   Novel behaviors have been predicted theoretically and observed in simulations and experiments, including  flocking~\cite{Vicsek1995}, large density fluctuations \cite{Ramaswamy2003,Narayan2007}, and spontaneous phase separation \cite{Tailleur2008,Fily2012,Redner2013}.
Walls and confined geometries are ubiquitous in realizations of active systems.  For example, sperm and bacteria often live near surfaces or in narrow channels, and these interfaces  strongly affect their dynamics~\cite{Lauga2006,Berke2008}.
Vibrated granular rods  spontaneously accumulate at the walls  even in the absence of hydrodynamic interactions~\cite{Kudrolli2008,Elgeti2009}.
Finally, mixtures of two types of active particles have been studied as minimal models of cell sorting in co-cultures and have been shown to segregate in bulk in the presence of adhesive interactions~\cite{Mehes2012,Julio2008,McCandlish2012}.

\begin{figure}
\includegraphics[width=1.00\columnwidth]{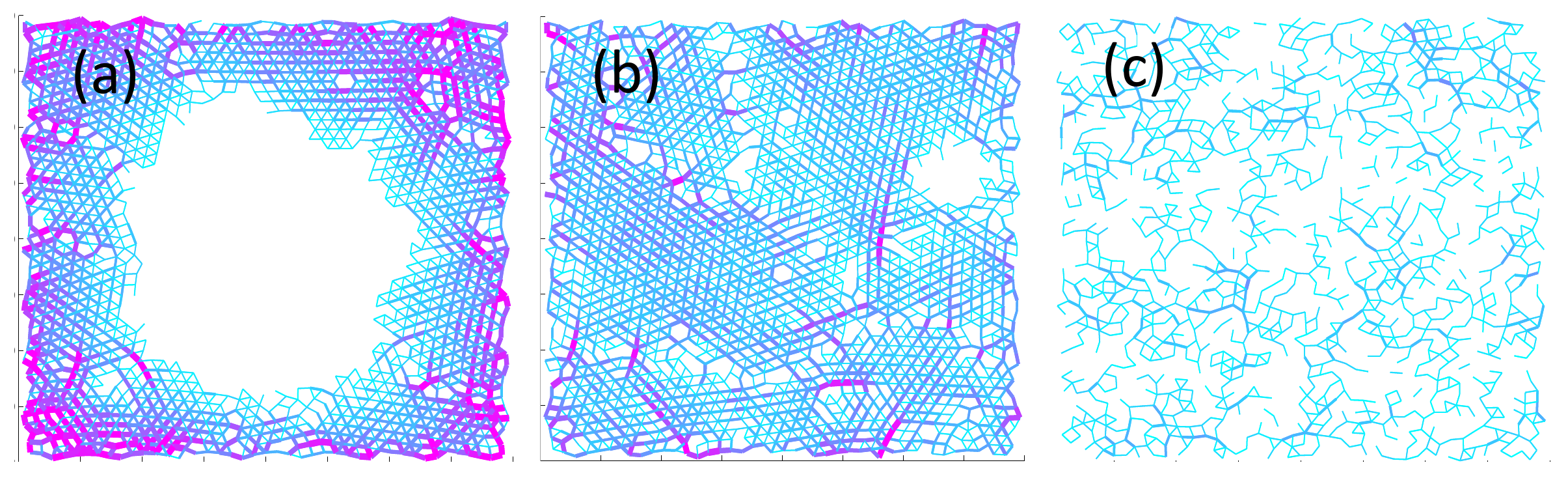}
\caption{(color online) Force chains at time $T=2000$ for $v_0=0.02$ displaying (a) aggregation at  $D_r=5\times10^{-5}$ and $\phi=0.672$, (b) jammed state at $D_r=5\times10^{-5}$ and $\phi=0.896$, and (c) homogeneous gas state at  $D_r=0.005$ and $\phi=0.672$. (Supplementary Movies 1-3)}
\label{fig:snapshots}
\end{figure}

In this paper we  study a minimal model of athermal self-propelled disks with purely repulsive interactions confined to a  box in two dimensions. Each disk performs a persistent random walk consisting of ballistic runs at speed $v_0$, randomized by rotational diffusion at rate $D_r$. We find that confined self-propelled particles aggregate at the walls provided their rotational diffusion is sufficiently slow (Fig.~\ref{fig:snapshots}). 
At low density, aggregation occurs when a particle travels ballistically across the container.  At high packing fraction $\phi$, however, a critical active speed $v_c(\phi)$ is required for wall aggregation even in the limit $D_r\rightarrow 0$. The onset of a nonzero value of $v_c$ in our active material correlates with the packing fraction at which non-active hard disks become ``jammed'' ~\cite{Donev2004}, i.e. exhibit a non-zero yield stress. The pressure of the active fluid, like the density, is spatially inhomogeneous as the particles seem to organize to optimally transmit stresses to the walls, as shown in Fig.~\ref{fig:snapshots}(a). As the jamming point is approached, the system becomes more uniform (Fig.~\ref{fig:snapshots}(b)) and the pressure begins to decrease with increasing density. This decrease occurs well below the jamming point and is associated with the onset of slow relaxation times due to strong caging effects that occur over a broad range of densities due to their activity. This non-monotonic dependence of pressure on density is unique to active systems. It is consistent with the non-monotonic dependence of pressure on temperature in a thermal active gas~\cite{Mallory2013} and on wall size in a dilute active gas~\cite{Ray2014}.  Finally,  this aggregation can be harnessed in a mixture of self-propelled particles of different sizes that segregates in the absence of any alignment or attraction (Fig.~\ref{fig:segregation}). The sense of the segregation (i.e., whether the large or the small disks accumulate on the outside) is determined by a  mean field calculation for the energy barrier generated by the repulsive interaction.  This segregation is reminiscent of cell sorting in embryonic development and is very different from the mechanisms that have been previously studied~\cite{Steinberg1963, Harris1976, Foty2005, Krieg2008, Manning2010}, which require differential cell adhesion or repulsion and postulate that cell sorting relaxes the tissue towards a free energy minimum, as in thermal systems.

\paragraph{ The Model.} We consider a system of $N$ monodisperse disks of radius $R$ in a square box of length $L$. The overdamped dynamics is  governed by Langevin equations for the position ${\bf r}_i$ of the center of the $i$-th disk and a unit vector ${\bf u}_i=\left(\cos\theta_i,\sin\theta_i\right)$ along the axis of self propulsion, 
\begin{equation}
\label{spp_eqn}
\partial_t\bm r_i=v_0\bm u_i+\mu\sum\limits_{j} \bm F_{ij}\;,~~~~~~~\partial_t\theta_i=\eta_i(t)\;,
\end{equation}
where $v_0$ is the active (self propulsion) speed and $\mu$ the mobility.  The particles interact via short-range repulsive forces $\bm F_{ij}$ proportional to the overlap between two disks, $\bm F_{ij}=k(2R-r_{ij}){\bf  \hat{r}}_{ij}$, with ${\bf r}_{ij}={\bf r}_i-{\bf r}_j={\bf \hat{r}}_{ij}r_{ij}$, and $k$ a force constant. The angular noise $\eta$ is white, with $<\eta_i(t)\eta_j(t')>=2D_r\delta_{ij}\delta(t-t')$ and $D_r$  the rotational diffusion rate.  Large immobile particles are glued to the walls of the box to implement the confinement.  At low density, each disk performs a persistent random walk and is diffusive at long times ($t\gg D_r^{-1}$), with an effective diffusion constant $D_a=v_0^2/2D_r$~\cite{Fily2012}. We treat $D_r$ as an independent parameter because in many realizations, including bacterial suspensions~\cite{Berg2004} and  active colloids~\cite{Palacci2013}, the rotational noise is athermal.  In these systems, $D_a$ is also typically two orders of magnitude larger than the thermal diffusivity, and so we neglect thermal noise in Eq.~\eqref{spp_eqn}.


Lengths and times are in units of the particle radius $R$ and the elastic time $\mu k$. Unless otherwise noted, the size of the box is $L=83$. Particle positions are initialized with a uniform random distribution inside the box, and orientations are random over the interval $[ 0, 2 \pi]$.   Equations~\eqref{spp_eqn} are integrated numerically using an Runge-Kutta algorithm for $t=9000$ timesteps.  This time interval is sufficient to ensure that the density profile of the system has reached steady state.
We explore the behavior of the system by varying the active velocity $v_0$, the rotational diffusion rate $D_r$, and the packing fraction $\phi=N\pi R^2/L^2$~\footnote{The values of packing fraction quoted below and in all figures have been adjusted to take into account the area occupied by the particles glued to the walls.}.

\paragraph{ Wall Aggregation.} 
\begin{figure}
\includegraphics[width=1.00\columnwidth]{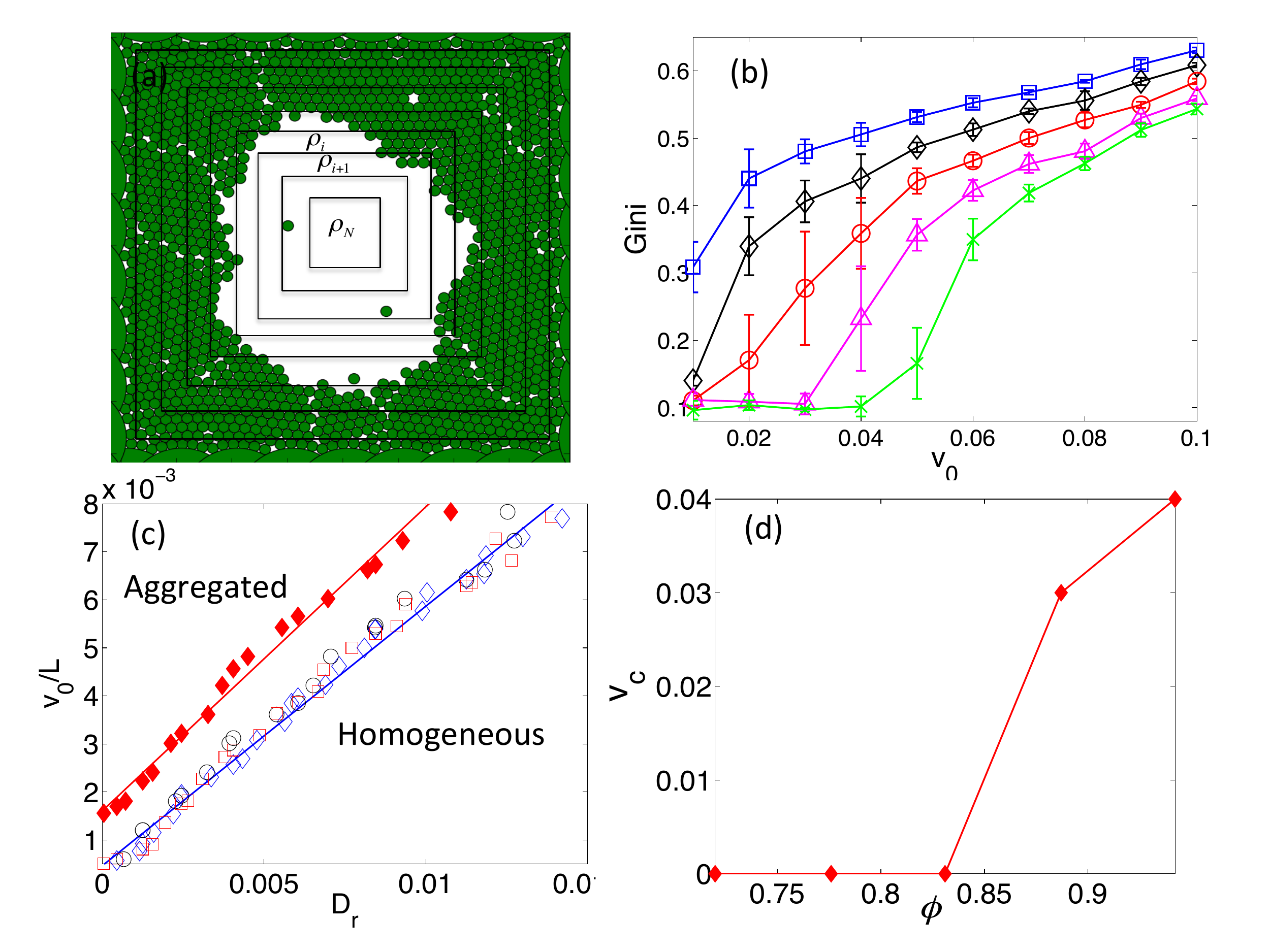}
\caption{(color online) (a) Diagram of nested square strips. (b) Gini coefficient vs. $v_0$ at various packing fractions $\phi=0.720, 0.776, 0.831, 0.887$ and $0.942$ (squares, diamonds, circles, triangles and crosses). The rotational noise is  $D_r=5\times10^{-5}$ and the total simulation time is $T=15000$. (c) Phase boundaries separating aggregated and homogeneous states in the plane of $v_0/L$ vs $D_r$. The open symbols are for  $\phi=0.40$ and $L=83,110,130$ (circles, squares, diamonds). The straight line is a fit to that data with $v_0=AL D_r$, where $A=0.5402$. Filled diamonds are for  $\phi=1.00$ and $L=83$.  The total simulation time is $T=2000$. (d) Critical speed at $D_r\rightarrow 0$ vs. packing fraction.}
\label{fig:aggregation}
\end{figure}
To quantify wall aggregation and the resulting density inhomogeneities we divide the system in $n_\Delta$ nested square strips of thickness $\Delta$ (Fig.~\ref{fig:aggregation}(a)) and calculate the gini coefficient~\cite{Gini2012}, given by
$g=\frac{1}{2N^2|\bar{\rho}|}\sum\limits_{i}\sum\limits_{j}|\rho_i-\rho_j|$,
with $\bar{\rho}$ the mean density, $\rho_i$ the number density of particles in the $i$-th  strip, and $\Delta=2R$. The gini coefficient approaches $0$ when the density is homogeneous and  $1$ when all particles are at the wall. The boundary separating homogenous states from aggregated states where the particles accumulate at the walls is obtained by a linear fit to isosurfaces of the gini coefficient, and correspond to $g=0.5$, shown in Fig~\ref{fig:aggregation} (c) for different values of $\phi$. 
At low $\phi$, aggregation occurs when $D_r$ is small and particles travel ballistically across the container. The phase boundary is well-described by $v_0/L \propto D_r$, which is the solid line through the open circles  in Fig.~\ref{fig:aggregation} (c). At high $\phi$, a finite value $v_c(\phi)$ is required for wall aggregation even in the limit $D_r\rightarrow 0$, as shown by the solid line through the closed diamonds in Fig.~\ref{fig:aggregation} (c).  The dependence on $\phi$ is seen in Fig.~\ref{fig:aggregation} (b), where the gini coefficient immediately rises from its minimal value for $\phi < 0.83$, and only rises at a finite $v_c$ for $\phi > 0.88$. The critical $v_c$ as a function of $\phi$ is shown in Fig.~\ref{fig:aggregation}(d).  The onset of a finite threshold for aggregation at $\phi\simeq 0.88$ coincides with the jamming point for monodisperse, athermal non-active hard disks~\cite{Donev2004}, and is consistent with active jamming in a disordered landscape~\cite{Reichhardt2014}.
\begin{figure}[!h]
\includegraphics[width=1.00\columnwidth]{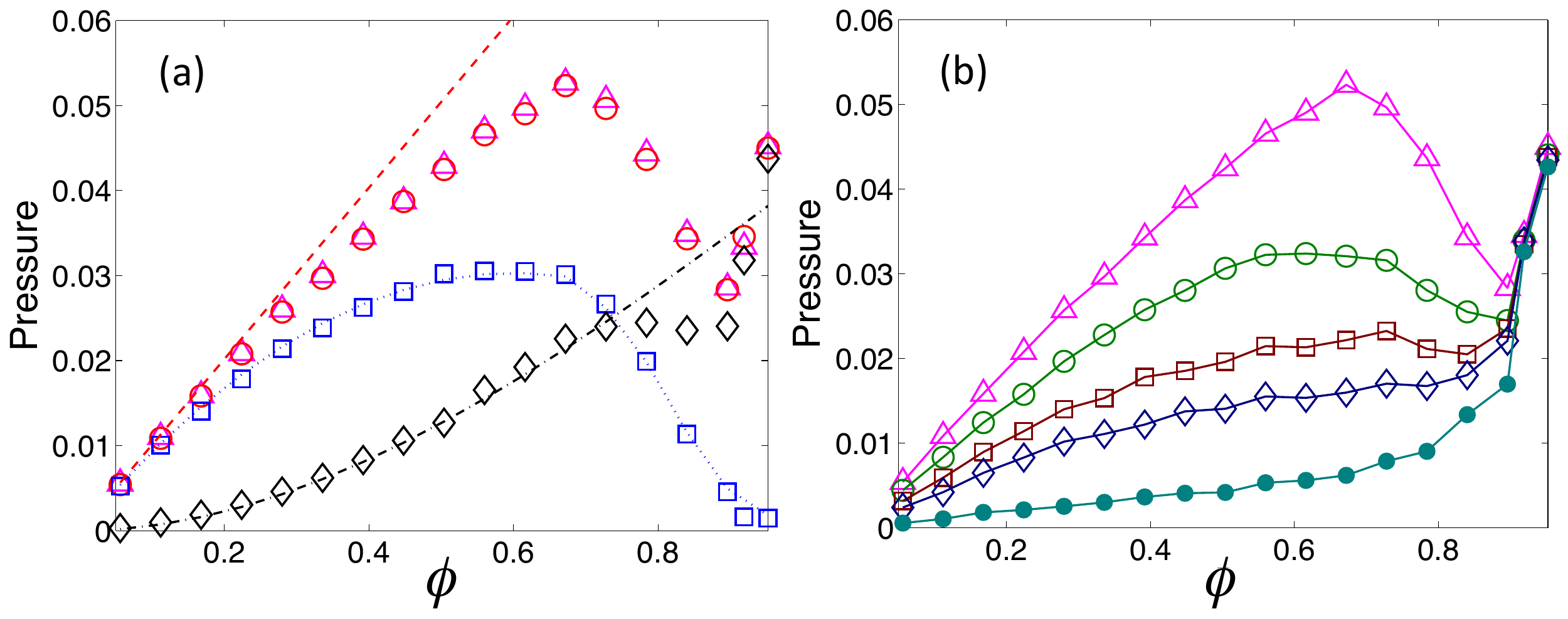}
\caption{(color online)  (a) Total pressure calculated from the IK formula (triangles) and as the force on the walls (circles) as a function of packing fraction at $v_0=0.02$ and $D_r=5\times10^{-5}$. The two calculations yield the same result. Also shown are the interaction (black diamond) and active (blue squares) contributions to the pressure. The dashed magenta line is the calculated  ideal gas  pressure with no fitting parameters. The black dot-dashed line is the calculated interaction pressure with $c=1.2$. (b) Total pressure for $D_r=5\times10^{-5},2\times10^{-4}, 4.5\times10^{-4}, 8\times10^{-4}$ and $5\times10^{-3}$(triangles, circles, squares, diamonds and filled circles).}
\label{fig:pressure}
\end{figure}
\paragraph{Pressure.} To quantify force distribution in our active fluid, we have evaluated the pressure both in the homogeneous and wall-aggregated states.  We define the pressure using the Irving-Kirkwood (IK) expression for the stress tensor given below ~\cite{IK1949}, augmented by a contribution from self-propulsion.  We have checked that this yields the same result as measuring the force per unit length on the walls of the container at all packing fractions.  The stress tensor $\sigma_{\alpha\beta}$ (with $\alpha,\beta=x,y$) is naturally separated in a contribution from interactions and an active contribution, as $\sigma_{\alpha\beta}=\sigma_{\alpha\beta}^{int}+\sigma_{\alpha\beta}^{a}$, with
\begin{equation}
\sigma_{\alpha\beta}^{int}=\frac{1}{L^2}\Big\langle\sum\limits_{i\neq j}F_{ij}^{\alpha}r_{ij}^{\beta}\Big\rangle,\hspace{0.05in}
\label{stress_defc}
\sigma_{\alpha\beta}^{a}=\frac{1}{L^2}\Big\langle\sum\limits_{i}F_{i,a}^{\alpha} r_i^{\beta}\Big\rangle,
\end{equation}
where ${\bf F}_{i,a}=(v_0/\mu){\bf u}_i$ is the active force on each disk. The pressure is the trace of the stress tensor,  $P=\sigma_{\alpha\alpha}/2=P_{int}+P_{a}$, shown in Fig.~\ref{fig:pressure}(a) as a function of $\phi$ for a small rotational diffusion rate $D_r=5\times10^{-5}$. For small $D_r$, where the system aggregates at the walls and exhibits strong density and pressure inhomogeneities (Fig.~\ref{fig:snapshots}(a)), the pressure is a strongly non-monotonic function of density and starts decreasing at $\phi\simeq 0.672$, well below jamming. At this packing fraction the density gradients start to smoothen, and the pressure becomes more homogeneous, as shown in Fig.~\ref{fig:force}(a), which displays the interaction force between particles as a function of distance to the wall. Fig.~\ref{fig:force}(b) shows the gini coefficients of density and pressure, demonstrating that the pressure inhomogeneity is a direct consequence of density inhomogeneity. Meanwhile, particles are caged by their neighbors. This  leads to ``self-trapping'', resulting in a suppression of their effective self-propulsion speed, as discussed in recent work on active phase separation~\cite{Tailleur2008,Fily2012,Fily2014,Redner2013,Cates2012}. In this region, although the system is fairly homogeneous, the transmission of force is impeded by crowding, resulting in an increased effective rotational diffusion rate and a sharp decrease in pressure. This description is supported by the correlation between the compressibility and homogeneity of the system, Fig~\ref{fig:force}(b).
The decrease in the forces that particles are able to transmit to the walls is most dramatic in the active pressure, that seems to essentially vanish near $\phi=0.907$, the packing fraction corresponding to perfect crystalline order in a triangular lattice. Fig.~\ref{fig:pressure}(b) shows that the pressure  non-monotonicity  diminishes with increasing $D_r$, and the system becomes thermal-like when $v_0/L\ll D_r$. In all cases, the pressure increases above  $\phi\simeq 0.88$ due to enforced overlap.
\begin{figure}[!h]
\includegraphics[width=1.00\columnwidth]{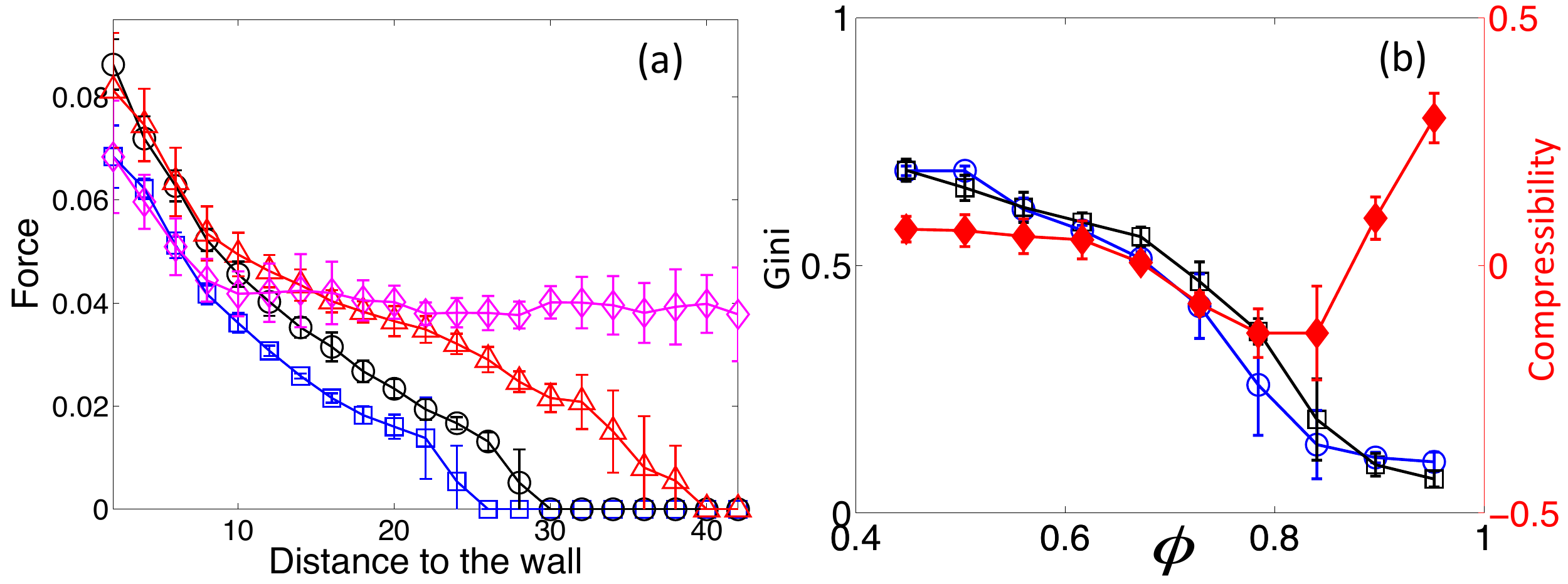}
\caption{(color online)  (a) Interaction force as a function of distance to the wall for packing fractions $\phi=0.56,0.672,0.784$ and $0.896$ (squares, diamonds, circles and triangles). (b) Gini coefficient of density (circles), gini coefficient of force (squares) and compressibility (filled diamonds) vs. packing fractions at $v_0=0.02$, $D_r=5\times10^{-5}$.}
\label{fig:force}
\end{figure}

The active pressure can be calculated analytically at low density  from the Langevin equations~\eqref{spp_eqn} neglecting interactions. The result corresponds to the pressure of an active ideal gas, also discussed in \cite{Mognetti2013}. Using $\left<u_{i\alpha}(t)u_{j\beta}(t')\right>=e^{-D_r|t-t'|}$,  we find $P_0(t)=\frac{\rho v_0^2}{2\mu D_r}\left(1-e^{-D_rt}\right)$ for the  ideal active gas pressure. 
In a container of side $L$, active particles eventually get stuck at the wall. For small, but finite $D_r$, we then define the ideal gas active pressure as $P_0=P_0(t=L/v_0)$, where $L/v_0$ is the time required by an active particle to travel ballistically through the container. The resulting expression $P_0=\frac{\rho v_0^2}{2\mu D_r}\left(1-e^{-D_rL/v_0}\right)$ interpolates between the thermal limit $P_0\approx \rho v_0^2/(s\mu D_r)$ for $D_r\gg v_0/L$ and the value $P_0\approx\rho v_0L/(2\mu)$ for $D_r\ll v_0/L$ corresponding to $N$ disks each exerting a uniform force $v_0/\mu$ on the walls. The ideal pressure of an active gas is shown in Fig.~\ref{fig:pressure}(a) as a dashed line and fits the data at low density. At high density, however, the active pressure shows a strong decrease with density indicative of strong caging and crowding that are not captured by the ideal gas calculation.  A fit to a mean-field theory that incorporates density-dependent velocity and rotational diffusion rates as discussed in the Supplementary Material is shown in Fig. 3(a) as a dotted line.

A simple expression for the interaction pressure can be obtained by modeling the system as concentric layers of particles aggregated at the walls and assuming that the particle overlap, hence the force that each layer exerts on the walls, increases linearly as the wall is approached. This estimate, described in the Supplementary Material, gives 
 $P_{int}=c\left(\frac{L v_0}{16 R}\phi^2-\frac{L v_0}{48 R}\phi^3\right)$, with $c$ a fitting parameter. A fit to this expression with $c=1.2$ is shown in Fig.~\ref{fig:pressure}.

\paragraph{Active Mixtures and Segregation.} The mechanisms responsible for athermal phase separation~\cite{Fily2012} and wall aggregation of purely repulsive self-propelled particles have remarkable consequences in mixtures. We simulate a binary mixture of small (S) and large (L)  self propelled particles with diameter ratio 1.4 to prevent crystallization.   Although  different in size, they interact via the same harmonic soft repulsive potential, with {\em equal} force constants $k_{SL} =k_{SS}=k_{LL}$, and with dynamics described by Eqs.~\eqref{spp_eqn}. The self-propulsion speeds are $v_S$ and $v_L$ respectively, and to reduce the number of parameters we have assumed equal mobilities for both types of particles.
\begin{figure}[!h]
\includegraphics[width=1.00\columnwidth]{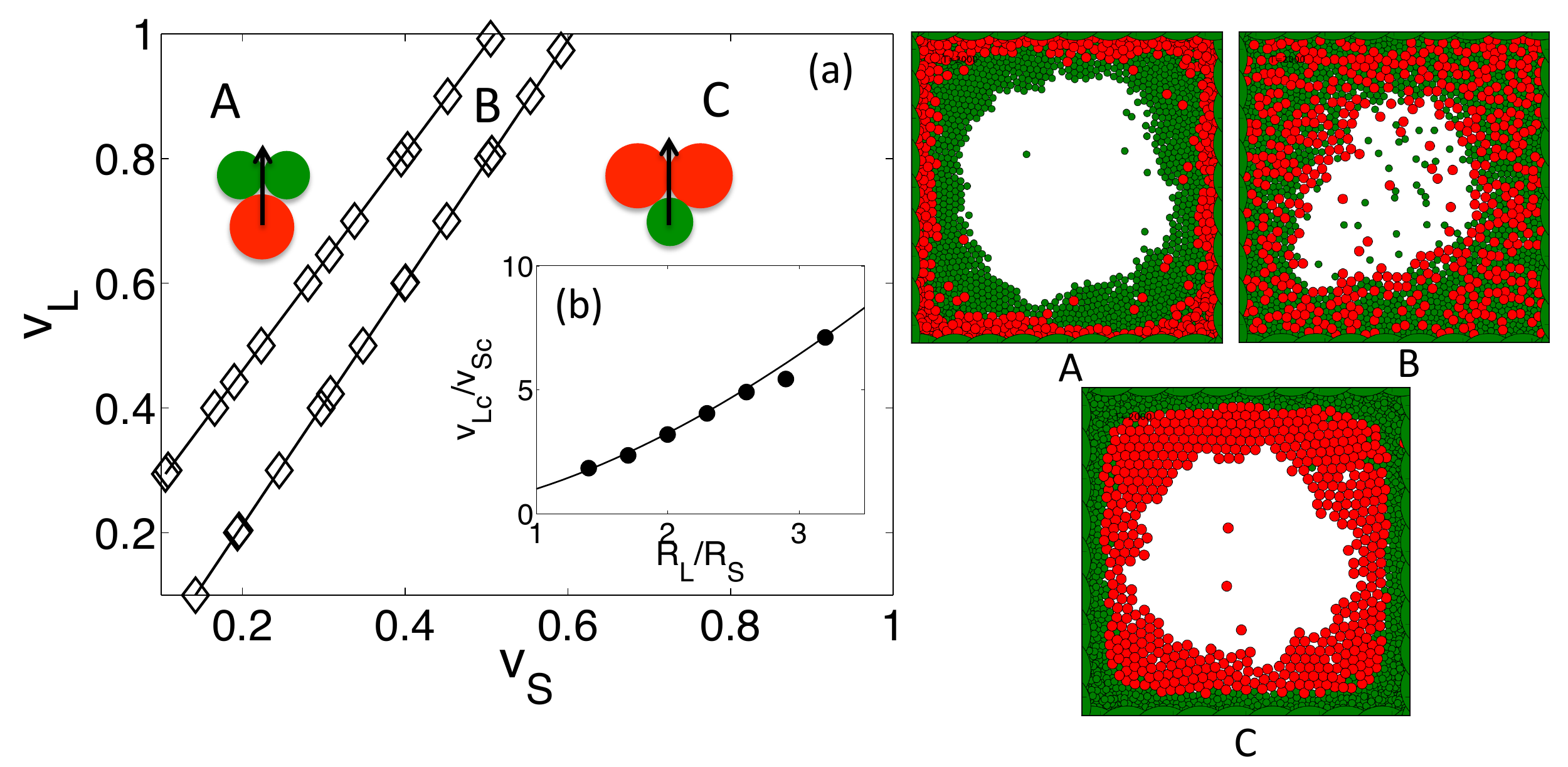}
\caption{(color online) Left: (a) Phase diagram showing the segregated and homogeneous states as functions of the active velocities   $v_S$ and $v_L$ (small particles are green and large ones are red online) for $D_r=5\times10^{-5}$ and a total packing fraction $\phi=0.9$, with each species occupying half of the packing fraction.  (b) Ratio $v_{Lc}/v_{Sc}$   as a function of the radii ratio $R_L/R_S$. The circles are from numerical simulation. The solid line is Eq.~\eqref{critical_velocity}. Right: snapshots of segregated and homogeneous states. The labels $A,B,C$ correspond to the states marked in the phase diagram. (Supplementary Movies 4-5)}
\label{fig:segregation}
\end{figure}

To quantify the spatial distribution of the two particle types, we define a segregation coefficient $S$:
\begin{equation}
\label{DS}
S=\frac{\sum\limits_i|\rho_{i}^L-\rho_{i}^S|}{\sum\limits_i \max[\rho_{i}^L,\rho_{i}^S]}
\end{equation}
where the shell width $\Delta$ is the large particle diameter and $\rho_i^{S,L}$ is the density of small/large particles in the $i$-th shell. With this definition, $S \rightarrow 0$ for a uniform distribution of L and S disks, and $S \rightarrow 1$ for complete segregation. 

When these purely repulsive disks are exactly the same except for their size ($v_S = v_L$), the system spontaneously segregates so that the small particles aggregate near the walls and the large particles are closer to the center of the box. We choose a critical value of $S=0.5$ to differentiate segregated state from mixed state.  

To better understand this suprising result, we study a phase diagram of the segregation as a function of the two self propulsion speeds $v_S$ and $v_L$, shown in Fig.~\ref{fig:segregation}. We find three distinct states: (A) a segregated state where all the large (red) disks have accumulated at the wall, with the small (green) ones closer to the center, (B) a mixed state where the particles have accumulated at the wall, but they are homogeneously distributed, hence $S \sim 0$ and (C) a segregated state where the small disks are near the walls and the large ones are near the center.  
The lower left hand corner of  Fig~\ref{fig:segregation} demonstrates that if both the small and large particle velocities are too small, the system remains mixed.  This suggests that particles must overcome a finite energy barrier in order to segregate. To quantify and test this assumption, we let $v_{Sc}$ ($v_{Lc}$) denote the critical velocity of the small (large) particles in the limit $v_{L} \rightarrow 0$ ($v_{S} \rightarrow 0$).
To estimate $v_{Sc}$, we derive an analytic expression for the velocity required for an active small particle to cross through two immobile large particles in contact with zero overlap, assuming that the small particle is moving directly perpendicular to the pair, as illustrated in Fig.~\ref{fig:segregation} (See Supplementary Material for details). This is a mean-field theory for energy barriers in a system exactly at the jamming transition.  We derive a similar expression for $v_{Lc}$, and calculate the ratio  $v_{Lc}/v_{Sc}$.  While the data in Fig.~\ref{fig:segregation}(a) are for a bidisperse mixture with diameter ratio 1.4, we calculate the velocity ratio as a function of the diameter ratio $x=R_L/R_S$, obtaining
\begin{equation}
\label{critical_velocity}
\frac{v_{Lc}}{v_{Sc}}=x^{-\frac{2}{3}}\frac{[1-(1+x)^{-2/3}]^{\frac{1}{2}}[(1+x)^{2/3}-1]}{[1-(1+\frac{1}{x})^{-2/3}]^{\frac{1}{2}}[(1+\frac{1}{x})^{2/3}-1]}
\end{equation}
This function $\frac{v_L}{v_S} (x)$ is plotted in Fig~\ref{fig:segregation}(b) as a solid line. We then extract numerical values of $v_{Lc}/v_{Sc}$ from the segregation boundary in simulations with different values of $R_L/R_S$. These numerical results are the data points in Fig ~\ref{fig:segregation} (b).  The remarkable overlap between the theory and simulation suggests that our mean field theory is valid and that asymmetric energy barriers for particles moving across one another are responsible for segregation.

\paragraph{Discussion.}
We have demonstrated that in the limit of small rotational noise, spherical self-propelled particles spontaneously accumulate at the walls of a container in the absence of any alignment or attractive interactions. At high density there is a finite threshold speed $v_c(\phi)$ for wall aggregation in the limit $D_r\rightarrow 0$. This speed vanishes at low density and becomes finite near the jamming transition, suggesting that the particles must overcome a finite yield stress to rearrange and accumulate at the walls.  The pressure displays a startling non-monotonic dependence on density.  When particles are aggregated at the walls the pressure increases with density, as the particles pack densely to optimize force transmission. Eventually, 
as the system approaches the jamming transition, both density and force distribution  become more homogeneous and the particles become caged by their neighbors,  losing the ability to self-organize to optimally transmit stress.  The net result is that the pressure decreases drastically with increasing density. We are currently implementing simulations  at constant pressure to interpret this surprising effect that could never happen in a thermal system.

In a mixture of active disks of two sizes we observe segregation in the absence of any adhesive interaction. 

This work was supported by the National Science Foundation through awards DMR-1305184 (MCM, XY), DGE-1068780 (MCM), BMMB-1334611(MLM), and DMR-1352184 (MLM), and by the Simons Foundation (MCM). MCM and XY also thank the KITP at the University of California Santa Barbara for hospitality during the final stages of the work. Finally we thank Silke Henkes and Yaouen Fily who developed the Brownian dynamics program used in the simulations and Jean-Francois Joanny for illuminating discussions.   

\section{Supplementary Material}

\begin{figure}[!h]
\begin{minipage}[t]{0.45\linewidth}
\centering
\includegraphics[width=\textwidth]{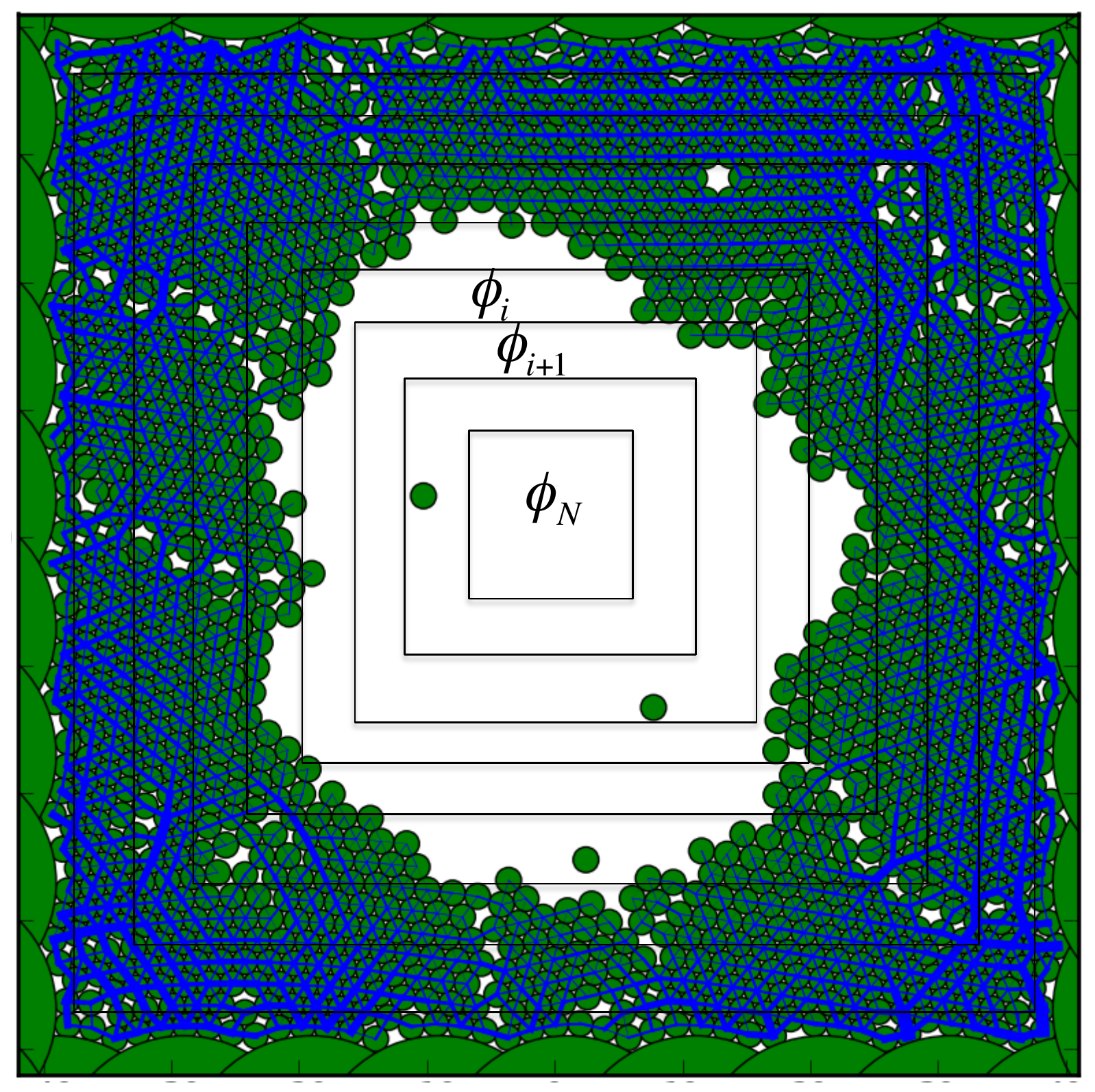}
\caption{(color online) Snapshot of an aggregated state with force chains (blue). The nested particle layers are also displayed. The overlap between particles increases as they approach the wall, indicating an inhomogeneous distribution of pressure, which is maximum  at the wall, as shown by the force chains.}
\label{fig:Snapshot}
\end{minipage}
\hfill
\begin{minipage}[t]{0.50\linewidth}
\centering
\includegraphics[width=\textwidth]{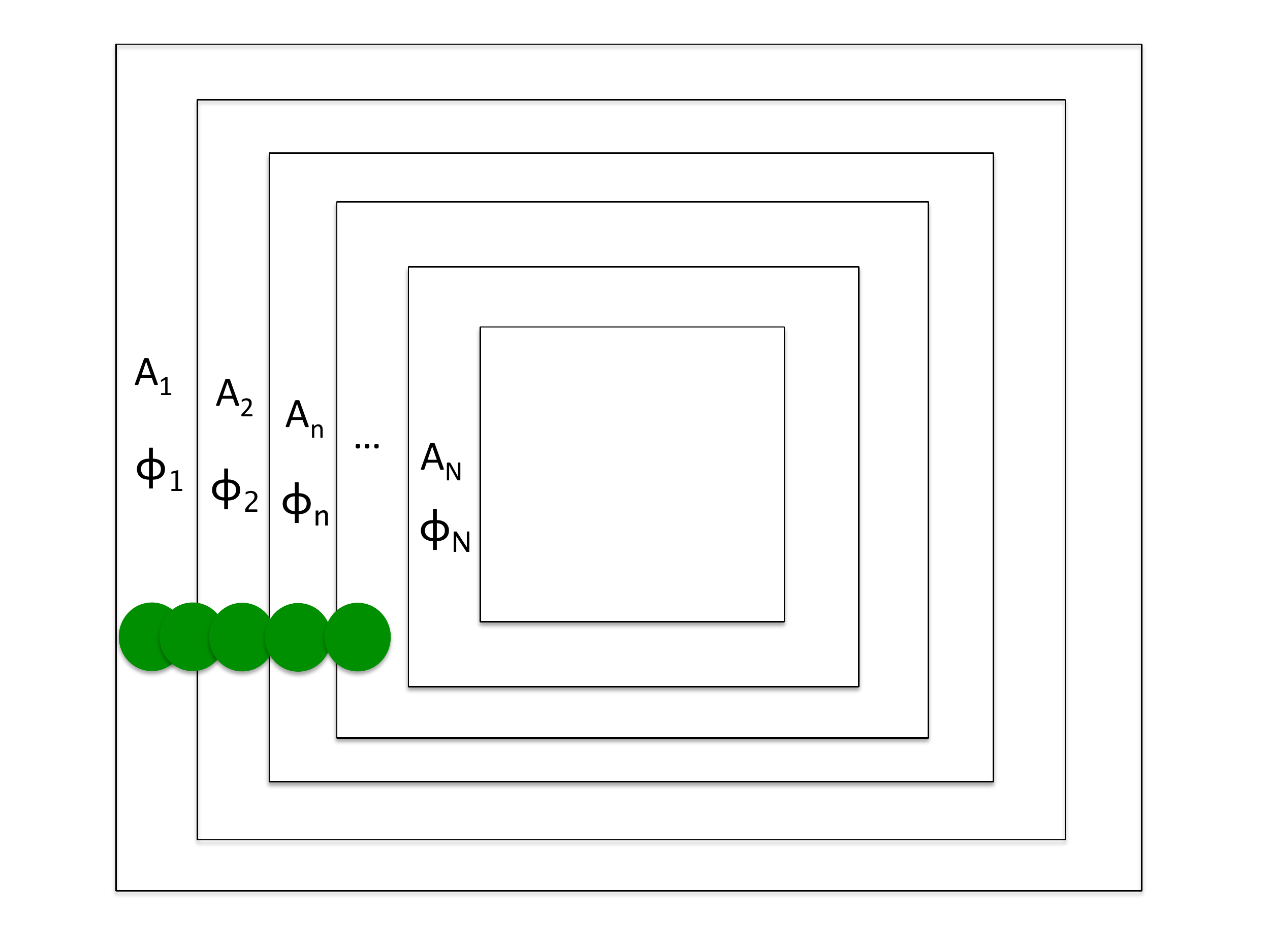}
\caption{(color online) The aggregated state is modeled  as a collection of  $N$ nested layers of particles, with a linear increase of overlap (or pressure) as the wall is approached. Each layer has area $A_n$ and is occupied by an area fraction $\phi_n$ of active particles.}
\label{fig:Model}
\end{minipage}
\end{figure}

\begin{figure}[!h]
\centering
\includegraphics[width=0.80\linewidth]{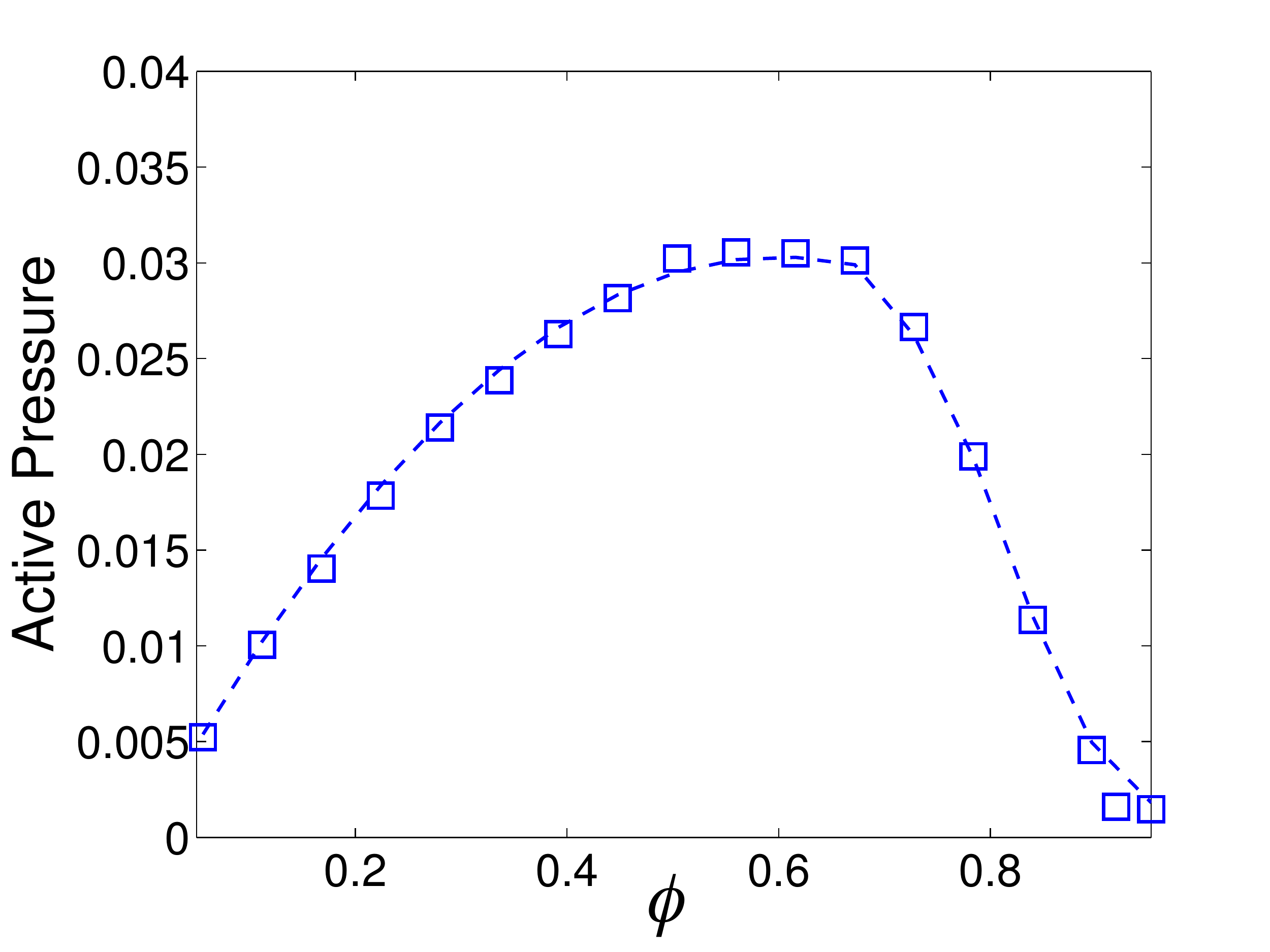}
\caption{Markers: Active pressure as a function of packing fraction for $v_0=0.02$ and $D_r=5\times10^{-5}$. Dashed line: Fitting using expression of ideal active gas pressure with density-dependent velocity and rotational diffusion rate.}
\label{fig:P_a}
\end{figure}

\begin{figure}[!h]
\centering
\includegraphics[width=1.00\linewidth]{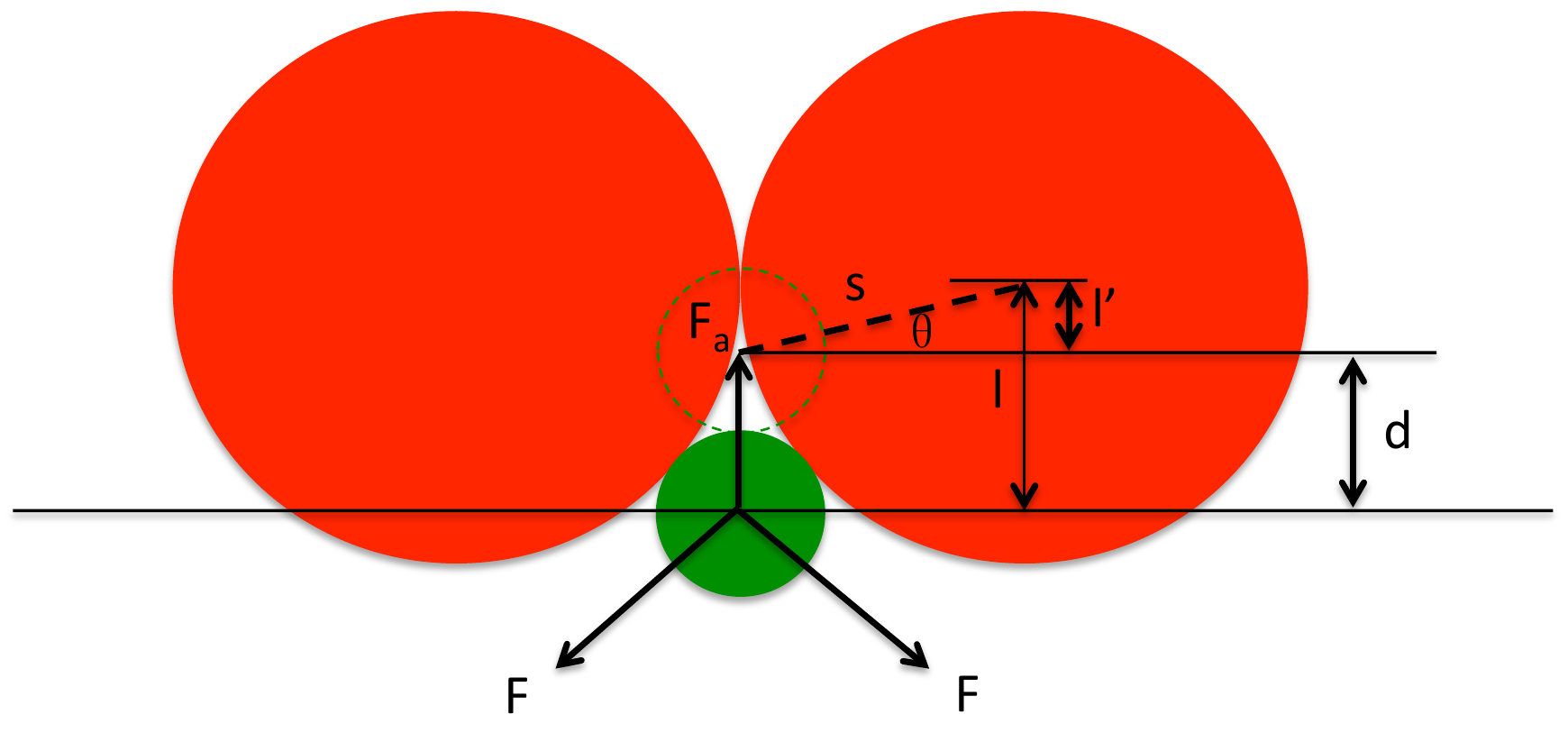}
\caption{Minimal model used to evaluate the barriers of segregation: a small active particle pushing its way through two adjacent, immobile large particles. $R_S$ and $R_L$ are the particles' radii. The small particle initially just  touches its neighbors, then travels a distance $d$ vertically with active velocity $F_a=v_0/\mu$. $F$ is the repulsive force between two particles. Other geometrical quantities are as labelled.}
\label{fig:diagram}
\end{figure}

\paragraph {\bf Interaction pressure of the aggregated state.}For simplicity, we consider a completely aggregated state, where the active force is balanced by the interaction force. We work in a coordinate system with axes along the principal direction of the stress tensor, and therefore drop the label of component for force and particle position. The trace of the stress is then given by
\begin{equation}
\label{stress_theory}
\sigma_{\alpha\alpha}=\frac{1}{L^2}\sum\limits_{i\neq j}F_{ij}r_{ij}\;,
\end{equation}
where the summation is over all interacting pairs. 
As illustrated in Fig ~\ref{fig:Model}, the interaction forces are transmitted through chains of particles, resulting in a larger interaction force/stress close to the wall. Given that our repulsive force is a linear function of overlap, we assume that the stress increases linearly as the wall is approached. This assumption is supported  by Fig. 4(a) in the paper. 
To proceed, we divide the system into $N$ nested particle layers, as shown in Fig ~\ref{fig:Model}. Each layer has the width of a particle diameter $2R$, area $A_n$ and occupies a fraction $\phi_n=A_n/L^2$ of the entire system's area. We assume that $\phi_n$  is also the packing fraction of particles in the $n$-th layer. Approximating the area of a layer as the sum of the area of four equal strips, we can write 
\begin{equation}
\label{area}
A_n=8LR-32R^2(n-1)\;.
\end{equation}
%
We assume that  the total packing fraction $\phi$ of the system is equal to the sum of $\phi_n$,
\begin{equation}
\label{packing_fraction}
\phi=\sum\limits_{n=1}^{N}\phi_n=\frac{1}{L^2}\left(8LRN+16NR^2-16R^2N^2\right)\;.
\end{equation}
Solving Eq.~\eqref{packing_fraction} for $N$ in terms of $\phi$, we obtain
\begin{equation}
\label{N}
N\simeq\frac{L}{8R}\phi+{\cal O}(R/L)\;.
\end{equation}
Now we proceed to calculate the stress. When the system is completely aggregated, the interaction forces are balanced by the active forces $F_a=v_0/\mu$.  Assuming that the force increases linearly as we approach the wall,  and imposing force balance between the interaction force $F_{ij}^n$ on particle $i$ in the $n$-th layer due to particle $j$ in the $n-1$ layer and the active forces, we can write
\begin{equation}
\label{force_balance}
F_{ij}^{n}=(n-1)F_a=(n-1)\frac{v_0}{\mu}\;.
\end{equation}
The  stress in the $n$th layer is then given by
\begin{equation}
\label{stress_layer}
\sigma_{\alpha\alpha}^{n}=\frac{1}{A_n}\sum\limits_{i\neq j}F_{ij}^{n}r_{ij}^{n}\;.
\end{equation}
Inserting Eq.~\eqref{force_balance}, we obtain
\begin{equation}
\label{stress_n_layer}
\sigma_{\alpha\alpha}^{n}=\frac{1}{A_n}\sum\limits_{1}^{CN_n/2}(n-1)\frac{v_0}{\mu}[2R-(n-1)\frac{v_0}{k\mu}]\;,
\end{equation}
where $C$ is a fitting parameter corresponding to the average contact number of a particle and $N_n$ is the number of particles in the $n$th layer. Expanding Eq.~\eqref{stress_n_layer} and keeping only terms to lowest order in $v_0$, we obtain
\begin{equation}
\label{stress_n_layer_final}
\sigma_{\alpha\alpha}^{n}=c\phi_n (n-1)v_0\;,
\end{equation}
where $c=\frac{C}{\pi R\mu}$ is a rescaled fitting parameter. Using $\phi_n=\frac{A_n}{L^2}$ and summing over the layers, we obtain an expression for  the total stress as
\begin{equation}
\begin{split}
\label{Total_stress}
\sigma_{\alpha\alpha}=\sum\limits_{n=1}^{N}\sigma_{\alpha\alpha}^{n}\simeq\int_{1}^{N}\sigma_{\alpha\alpha}^{n}(n)dn\\=\int_{1}^{N}\left[\frac{8Rcv_0}{L}(n-1)-\frac{32R^2cv_0}{L^2}(n-1)^2\right]dn
\end{split}
\end{equation}
where the sum over layers has been  replaced  by an integration.  Carrying out the integration we find
\begin{equation}
\begin{split}
\label{Total_stress}
\sigma_{\alpha\alpha}=\frac{4Rcv_0}{L}(N-1)^2-\frac{32R^2cv_0}{3L^2}(N-1)^3\\\simeq\frac{4Rcv_0}{L}N^2-\frac{32R^2cv_0}{3L^2}N^3=c\left(\frac{Lv_0}{16R}\phi^2-\frac{Lv_0}{48R}\phi^3\right)
\end{split}
\end{equation}
The pressure of the system is defined as
\begin{equation}
\label{Total_stress}
P=\frac{\sigma_{\alpha\alpha}}{2}
\end{equation}
To fit the data for  for $k=1$, $\mu=1$, $R=1$ and $L=80$ yields $c=1.2$, corresponding to an average contact number of 6.
\paragraph {\bf Non-monotonic Active Pressure.}The suppression of self-propulsion due to caging can be incorporated in a mean-field fashion by replacing the active speed $v_0$ in the ideal gas pressure by a density-dependent speed $v(\phi)$, as suggested by recent work on phase separation of active particles~\cite{Fily2012}. We also speculate that crowding effectively increases the rate of rotational diffusion as particles rattle around the confining cage and incorporate this effect into a density-dependent rotational diffusion rate, $D_r(\phi)$, that is enhanced at packing fraction above $\phi=0.672$,  where the active pressure starts to decrease. A fit to the active pressure using the ideal gas formula $P_a(t)=\frac{\rho v^2}{2\mu D_r}\left(1-e^{-D_rt}\right)$ with $v(\phi)=v_0(1-\lambda\phi)$ ~\cite{Fily2012} and $D_r(\phi)=\Theta(\phi-\phi_c)\exp[\alpha(\phi-\phi_c)]$, where $\Theta(\phi-\phi_c)$ is the Heaviside step function and $\phi_c=0.672$, is shown in Fig.~\ref{fig:P_a} as a dashed line. The fitting parameters are $\lambda=0.8$ and $\alpha=13$. 
\paragraph{\bf Segregation Barriers.}To evaluate the barrier that particles must overcome for segregation, we consider the geometry shown in Fig.~\ref{fig:diagram}  showing a small active particle of radius $R_S$ that has to make its way through two immobile large particles of radius $R_L$. 
For the small particle to travel through the barrier imposed by the two large ones, the active force $v_0/\mu$ has to overcome the maximum of the repulsive force $F_{rep}$.  This defines  a critical active velocity $v_{Sc}$ for the small particle. 
To calculate this barrier  we assume that the small active particle initially just touches its neighbors, then travels upward  a distance of $d$. At this point, the net repulsive force is 
\begin{equation}
F_{rep}=2Fsin\theta\;,
\end{equation}
where $F=k[(R_L+R_S)-\sqrt{(l-d)^2+R_L^2}]$ is the repulsive force between two particles and $sin\theta=l'/s$. Geometrical considerations lead  $l'=l-d$, $s=\sqrt{(l-d)^2+R_L^2}$ and $l=\sqrt{r^2+2R_SR_L}$, allowing us to express  $F_{rep}$ solely in terms of $d$.

The critical active velocity $v_{Sc}$ is defined as the maximum value of $\mu F_{rep}(d)$. This gives 
\begin{equation}
\begin{split}
v_{Sc}=2\mu k [(R_L+R_S)R_L^2]^{1/3}[1-(1+\frac{R_S}{R_L})^{-2/3}]^{1/2}\\\times[(1+\frac{R_s}{R_L})^{2/3}-1]\;.
\label{eqn:critical}
\end{split}
\end{equation}
This is the critical velocity of an active particle with radius $R_S$ pushing through two immobile particles of radius $R_L$. The critical velocity for the reversed configuration, corresponding to a particle of radius $R_L$ pushing through two particles  of radius $R_S$, can be obtained by interchanging $R_S$ and $R_L$. Segregation occurs when either species has an active velocity above the critical value. Note that particles with different radii have different critical velocities. If, for instance,  $R_S<R_L$, then $v_{Sc}<v_{Lc}$, and small particles  will aggregate to the outside, next to the wall,   when both species have the same active velocity. 

Finally,  the ratio of active velocities of the two species can be written as a function of their radii ratio as
\begin{equation}
\label{critical_velocity}
\frac{v_{Lc}}{v_{Sc}}=x^{-\frac{2}{3}}\frac{[1-(1+x)^{-2/3}]^{\frac{1}{2}}[(1+x)^{2/3}-1]}{[1-(1+\frac{1}{x})^{-2/3}]^{\frac{1}{2}}[(1+\frac{1}{x})^{2/3}-1]}\;,
\end{equation}
where $x=R_L/R_S$. This result is compared with the numerical result in Fig.5(b) of the paper. The excellent agreement supports our simple model.
\section{Movie Captions}
\begin{flushleft}
{\bf Supplementary Movies 1-3}
\end{flushleft}
These three movies display the aggregated, jammed and homogeneous gas states of a monodisperse system for $v_0=0.02$. Large immobile particles are glued to the  wall to confine the system. The blue lines describe the direction of the interaction forces between active particles. The thickness of the lines is proportional to the strength of force.

\paragraph{\bf Supplementary Movie 1.}

Aggregated state at $\phi=0.672$ and $D_r=5\times10^{-5}$. Active particles travel ballistically across the box and aggregate at the walls, leaving a void at the center. The inhomogeneous density leads to  force gradients, with the interaction force decaying away from the wall.

\paragraph{\bf Supplementary Movie 2.}

Jammed state at $\phi=0.896$ and $D_r=5\times10^{-5}$. Motion of the active particles is suppressed by local caging, leading to  crystalline domains separated by grain boundaries. Rattling of the particles in their cages promote local self-organization and the density and interaction forces remain globally homogeneous.

\paragraph{\bf Supplementary Movie 3.}

Homogeneous gas state at $\phi=0.672$ and $D_r=5\times10^{-3}$. For this  large value of the rotational diffusion rate  the system behave like a thermal gas with a homogeneous density and force distribution.

\begin{flushleft}
{\bf Supplementary Movies 4-5}
\end{flushleft}
These two movies display spontaneous segregation in a mixture of non-adhesive active particles with different sizes or activities. The total packing fraction is $\phi=0.90$, with both species occupying half of the space. The rotational diffusion rate $D_r=5\times10^{-5}$ and the radii ratio is $1:1.4$.

\paragraph{\bf Supplementary Movie 4.}

Spontaneous segregation at $v_S=v_L=0.3$, where the small particles are closer to the wall.

\paragraph{\bf Supplementary Movie 5.}

Spontaneous segregation at $v_S=0.1$ and $v_L=0.4$, where the large particles are closer to the wall.

\end{document}